# Resonance Features of the Conductance of Open Billiards with the Spin–Orbit Interaction[1]

Alexander I. Malyshev, Galina G. Isupova

Lobachevsky State University of Nizhni Novgorod, Nizhni Novgorod, Russia
e-mail: malyshev@phys.unn.ru

The transport properties of a circular billiard with attached channels, which is an open system, have been studied in the presence of the Dresselhaus and Rashba spin–orbit interactions. It has been shown that this interaction leads to the appearance of additional Fano resonances in the energy dependence of the conductance, the width of which is proportional to the fourth power of the spin–orbit coupling constant.

It is known that, in solid-state physics, the spin–orbit interaction plays a fundamental role, because it determines the electron quantum states and leads to multiple transport and optical effects, many of which are of applied interest [1, 2].

One of these new effects is discussed in this work. Let us consider an open quantum billiard with attached input and output channels with the spin–orbit interaction present both inside the billiard and in the channels (see Fig. 1). Although the type of the spin–orbit interaction is not important when considering the effect of interest, for definiteness, we accept the Dresselhaus spin–orbit interaction [3]

$$\hat{H}_D = \frac{\beta}{\hbar}\left(\hat{\sigma}_x \hat{p}_x - \hat{\sigma}_y \hat{p}_y\right). \quad (1)$$

Comments concerning the Rashba spin-orbit interaction [4] will be made when necessary.

First, it is necessary to determine the structure of the wavefunction both inside the billiard and in the attached channels.

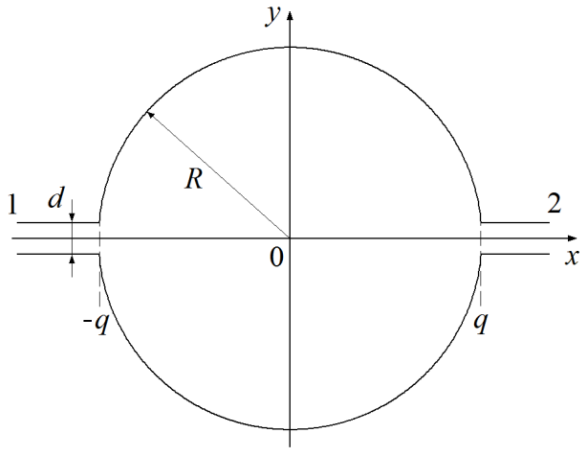

**Fig. 1.** Open billiard with the input (1) and output (2) channels.

The energy spectrum of a free particle with the wave vector $\vec{k}$ in the presence of the Dresselhaus spin–orbit interaction is as follows:

$$E_\lambda(k) = \frac{\hbar^2 k^2}{2m} + \lambda \beta k, \quad (2)$$

and its wavefunction is the spinor

$$\psi_{\lambda,\vec{k}} = \frac{e^{i\vec{k}\vec{r}}}{\sqrt{2}}\begin{pmatrix} 1 \\ \lambda e^{i\phi(\vec{k})} \end{pmatrix}. \quad (3)$$

Here, $\phi(\vec{k}) = \arg(k_x - ik_y)$ and $\lambda = \pm 1$ is the quantum number that corresponds to the two surfaces of the dispersion law split by the spin-orbit interaction (see, e.g., [5]). The states corresponding to constant energy are located in the $(k_x, k_y)$ plane at the circles with the radii

$$k_\pm = \frac{1}{\hbar}\left(\sqrt{2mE + \frac{m^2\beta^2}{\hbar^2}} \mp \frac{m\beta}{\hbar}\right), \quad (4)$$

where the subscripts "±" indicate the sign of the corresponding quantum number λ.

To characterize the states in the channels attached to the billiard, we consider a two-dimensional electron gas in an infinite channel in the presence of the Dresselhaus spin–orbit interaction. In this case, the Hamiltonian has the form

$$\hat{H} = \frac{\hat{p}_x^2 + \hat{p}_y^2}{2m} + \hat{H}_D + V(y), \quad (5)$$

where the potential $V(y)$ describes an infinitely deep potential well as follows:

$$V(y) = \begin{cases} 0 & \text{for } |y| < d/2, \\ \infty & \text{for } |y| \geq d/2. \end{cases} \quad (6)$$

Since $p_x$ is an integral of motion in this system, the solution of the stationary Schrödinger equation can be represented in the form

$$\psi = \frac{e^{ik_x x}}{\sqrt{2}}\begin{pmatrix} a(y) \\ b(y) \end{pmatrix}, \quad (7)$$

where the functions $a(y)$ and $b(y)$ satisfy the zero boundary conditions at the channel walls. It is convenient to expand these functions in a series of eigenfunctions of the transverse modes in the channel without the spin–orbit interaction $\varphi_n(y) = \sqrt{2/d}\sin[\pi n(y+d/2)/d]$, which provides the automatic fulfillment of the boundary conditions (see, e.g., [6]).

Before analyzing the calculation data, it is necessary to specify the units of measurement. Let the dimensionless Planck constant and the effective mass of carriers be equal to unity. The energy and spin–orbit coupling constant units are represented in terms of the length unit $l_0$ as $e_0 = \hbar^2/ml_0^2$ and $\beta_0 = \hbar^2/ml_0$, respectively. Below, dimensionless quantities are marked wavy symbols.

An example of the energy spectrum in the infinite channel with the Dresselhaus spin–orbit interaction normalized for convenience by the energy of the first transverse mode $E_1 = \pi^2\hbar^2/2md^2$ is given in Fig. 2. The spectrum consists of a series of branches split by the spin–orbit interaction. In the lower pair of branches, four wave states with the vectors $\pm k_1$ and $\pm k_2$ propagating to the right and left along the channel correspond to the fixed energy. A pair of states with the wave vectors $k_1$ and $-k_2$ differ from a pair of states with the wave vectors $k_2$ and $-k_1$ in the spin polarization, i.e., the components of the spin density $\bar{s}_i(x,y) = (\hbar/2)\psi^+\hat{\sigma}_i\psi$ for each point of the $(x,y)$ space have the opposite signs. Note that $\bar{s}_y(x,y) \equiv 0$, since the $a(y)$ and $b(y)$ functions here can be chosen to be real.

The solution of the Schrödinger equation in the channel with the Rashba spin–orbit interaction has a similar structure. However, in this case, the purely imaginary spinor component $a(y)$ corresponds to the real component $b(y)$ and vice versa; as a result, the $x$ component of the spin density is zero, i.e., $\bar{s}_x(x,y) \equiv 0$.

Let us now calculate the wavefunction in the structure with the circular billiard (see Fig. 1). It is necessary to solve the stationary Schrödinger equation with the Hamiltonian

$$\hat{H} = \frac{\hat{p}_x^2 + \hat{p}_y^2}{2m} + \hat{H}_D + V(x,y), \quad (8)$$

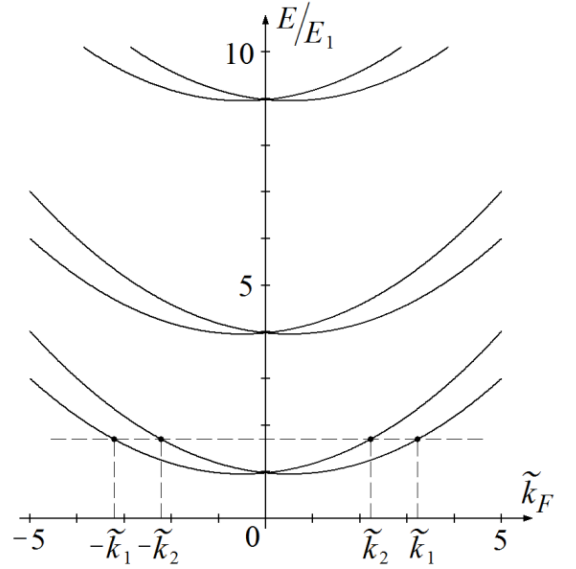

**Fig. 2.** Fragment of the energy spectrum of the electron in the quasi-one-dimensional channel with the Dresselhaus spin–orbit interaction at $\tilde{\beta} = 0.5$ and $\tilde{d} = 1$.

where $V(x,y)$ describes the infinite jump of the potential at the billiard boundary and the walls of the attached channels.

We consider the first pair of the spectral branches as follows (see Fig. 2). Let a wave with the wave vector $k_1$ enter channel 1 (on the left) and the transmitted waves with the vectors $k_1$ and $k_2$ and the amplitudes $c_1$ and $c_2$, respectively, propagate in channel 2 (on the right). The reflected waves with the wave vectors $-k_1$ and $-k_2$ and the amplitudes $c_3$ and $c_4$, respectively, also propagate in channel 1. Thus, the wavefunction has the form

$$\psi_{out}^{(1)}(x,y) = \frac{e^{ik_1 x}}{\sqrt{2}}\begin{pmatrix} a_1(y) \\ b_1(y) \end{pmatrix} +$$
$$+ c_3 \frac{e^{ik_1 x}}{\sqrt{2}}\begin{pmatrix} a_3(y) \\ b_3(y) \end{pmatrix} + c_4 \frac{e^{-ik_2 x}}{\sqrt{2}}\begin{pmatrix} a_4(y) \\ b_4(y) \end{pmatrix} \quad (9)$$

in the input channel and the form

$$\psi_{out}^{(2)}(x,y) = c_1\frac{e^{ik_1 x}}{\sqrt{2}}\begin{pmatrix} a_1(y) \\ b_1(y) \end{pmatrix} + c_2\frac{e^{ik_2 x}}{\sqrt{2}}\begin{pmatrix} a_2(y) \\ b_2(y) \end{pmatrix} \quad (10)$$

in the output channel. Here, $q = \sqrt{R^2 - d^2/4}$. Let us write the solution in the internal region of the billiard as a superposition of plane waves:

$$\psi_{in}(x,y) = \frac{1}{\sqrt{q}}\int_0^{2\pi}\left[c(\theta)e^{ik_+(x\cos\theta+y\sin\theta)}\begin{pmatrix} 1 \\ e^{-i\theta} \end{pmatrix} +\right.$$

$$+ d(\theta)e^{ik_-(x\cos\theta+y\sin\theta)}\begin{pmatrix}1\\-e^{-i\theta}\end{pmatrix}\Bigg]d\theta \qquad (11)$$

where $\theta$ is the angle in the $(k_x, k_y)$ plane measured from the positive direction of the $k_x$ axis and $k_\pm$ are determined in Eq. (4). To take into account the Rashba spin–orbit interaction, it is necessary to replace $\exp(-i\theta)$ by $-i\exp(i\theta)$ in Eq. (11), the quantities $k_\pm$ are determined analogously.

Then, we join the solutions of Eqs. (9)–(11) and require that Eq. (11) satisfies the zero boundary conditions on the billiard walls. All these requirements can be provided by the application of the method that was proposed in [7] and was expanded in [8] to the case of the two-component wavefunction. As a result, it is possible to analyze the features of the distribution of the probability density and the components of the spin density and to calculate the conductance. In this case, the latter can be found from the Landauer formula

$$G = \frac{e^2}{h}\left(|c_1|^2 + |c_2|^2\right). \qquad (12)$$

An example of the conductance dependence on $k_F$ characterizing the total energy, i.e., the Fermi energy $E_F = \hbar^2 k_F^2/2m$, in the system without the spin–orbit interaction and with the Dresselhaus spin–orbit interaction is shown in Fig. 3.

An interesting fact is that the amplitude $c_2$ does not contribute to the conductance; its value did not exceed $10^{-5}$ in all experiments. The same refers to the amplitude $c_3$. Thus, the spin polarization of waves transmitted and reflected in the studied billiard does not change. This agrees with the results reported in [9], where the conservation of the spin polarization was proved analytically for an arbitrary billiard with the Rashba spin–orbit interaction. It is extremely important that waves of both polarizations equally propagate inside the billiard: terms with $c(\theta)$ and $d(\theta)$ in Eq. (11) are of the same order of magnitude.

Although the problem of the convergence of the used method requires a fairly delicate approach, the method itself allows one to observe quite fine effects. For example, it is easy to note (see Fig. 3) that the inclusion of the spin–orbit interaction term into the Hamiltonian leads to the appearance of additional Fano-type resonances in the dependence of the conductance on $k_F$ (see review [10] and references therein). One can see six such resonances in Fig. 3b, which can be easily combined in pairs. Inside these pairs, the zeros of the resonances are oriented toward each other (see inset in Fig. 3b), although single resonances also occur at higher energies. It is remarkable that an increase in the spin–orbit coupling constant mainly leads to the broadening of these resonances and almost does not affect the positions of the peaks. This is illustrated in Fig. 4, where the $k_F$ positions of units ("1") and zeros ("0") for the resonances marked as 1 and 2 in

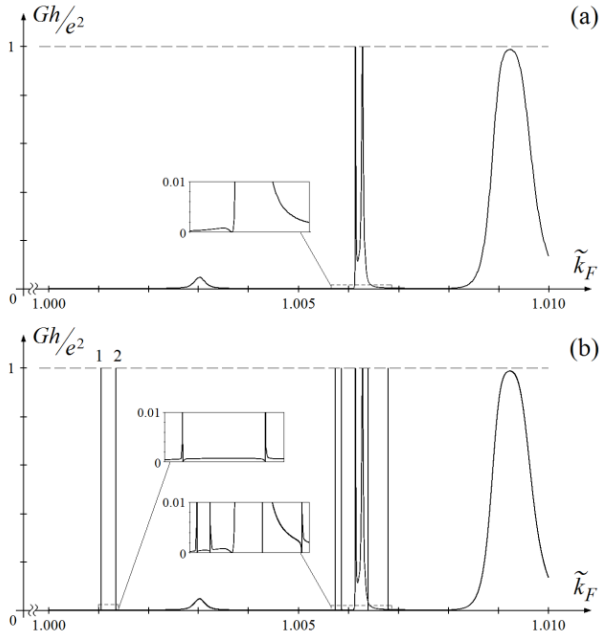

**Fig. 3.** Fragment of the $k_F$ dependence of the conductance of the system with the billiard (a) without spin-orbit interaction and (b) with Dresselhaus spin-orbit interaction at $\tilde{\beta}=0.003$ for $\tilde{d}=1$ and $\tilde{R}=15$. The enlarged fragments of the plots are shown in the inset.

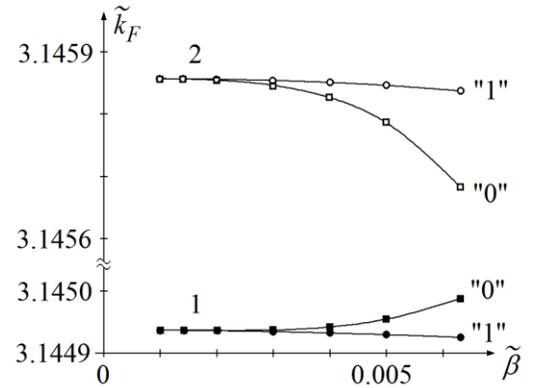

**Fig. 4.** Position of units and zeros for resonances *1* and *2* marked in Fig. 3b versus the Dresselhaus spin–orbit coupling constant.

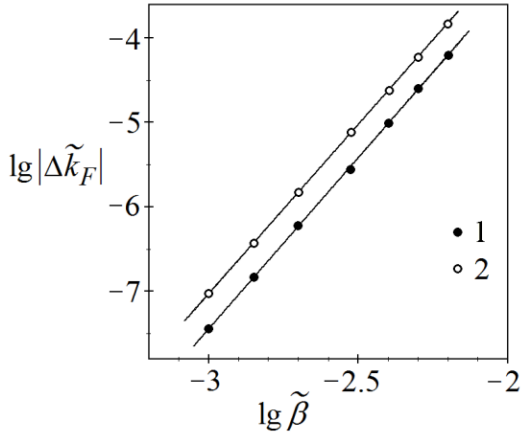

**Fig. 5.** Widths of resonances *1* and *2* versus the Dresselhaus spin–orbit coupling constant. The slopes of the straight lines plotted by means of the least-squares method are 4.03 and 3.99, respectively.

Fig. 3b are shown. It can be seen in the figure that the positions of the peaks change very slightly compared to the shifts in the zeros of the resonances.

The plot given in Fig. 5 is more representative. Here, the dependence of the width of these two resonances (the distances $\Delta k_F$ between the unit and zero) on the parameter of the Dresselhaus spin–orbit interaction is shown in the log–log scale. The slopes of both approximating lines are four with an accuracy of several hundredths. Therefore, the width in $k_F$ and energy (since $\Delta E_F \approx \hbar^2 k_F \Delta k_F / m$) of the Fano resonances due to the addition of the spin–orbit interaction in the studied system is proportional to the fourth power (!) of the spin–orbit coupling constant. The usage of the Rashba spin–orbit interaction in the calculations gives the same result.

Apparently, in this situation, there is every reason to say that the Fano resonances collapse [11, 12] in this case when the spin–orbit coupling constant tends to zero. At the same time, this allows one to state that an arbitrarily weak spin–orbit interaction has a noticeable effect at some energy values (in the extremely narrow regions) leading to the resonance features of the conductance. Undoubtedly, the further problem should be to clearly establish the origin of the appearances of the additional resonances and to establish the exact interrelation between their position, width, and the spin–orbit coupling constant.

In summary, let us give some numerical estimates. If the width of the input and output channels is chosen to be 30 nm, then the billiard diameter is 0.9 μm. Using a value of $0.067 m_e$ for the effective mass of the conduction electrons in GaAs, we obtain a measurement unit of 38 meV·nm for the spin–orbit coupling constant. In this situation, the interval of the spin–orbit coupling constants from 0 to $0.006 β_0$ taken for plotting the curves in Figs. 4 and 5 is less than the typical range of the β values in the real structures. Note again that all results discussed here hold also for the Rashba spin–orbit interaction, moreover, at the same numerical values of the α parameter as it was taken for the Dresselhaus spin–orbit coupling constant.

We are grateful to A.M. Satanin for fruitful discussions of the physical and mathematical aspects of the present problem. This work was supported by the Russian Foundation for Basic Research (project no. 09-01-00268) and the Dynasty Foundation.

---

[1] It's a slightly corrected version of the paper published in JETP Letters Vol. 94, Issue 7, pp. 556-559 (2011). Some new results will be available soon in Bulletin of the Russian Academy of Sciences: Physics Vol. 77, No. 1, pp. 78–82 (2013).